\documentclass[aps,prl,preprint,tightenlines,superscriptaddress]{revtex4-2}

\usepackage{graphicx} 
\usepackage{dcolumn}  
\usepackage{multirow}
\usepackage{relsize}
\usepackage{physics}
\usepackage{xcolor}
\graphicspath{{ps}}

\renewcommand{\arraystretch}{1.1}

\newcommand{\ra}{\rightarrow}
\newcommand{\mevcc}{~{\rm MeV}/c^2}
\newcommand{\mev}{~{\rm MeV}}
\newcommand{\Xic}{\Xi_{c}(2970)^+}
\newcommand{\xic}{\Xi_{c}}
\newcommand{\xics}{\Xi_{c}(2645)^0}
\newcommand{\xicp}{\Xi_{c}^{\prime0}}
\newcommand{\Xicdecay}{ \Xi_{c}(2970)^+ \ra \Xi_c(2645)^{0}\pi^{+}/\Xi_c^{\prime0}\pi^{+} }
\newcommand{\xicspi}{ \Xi_c(2645)^{0}\pi^{+} }
\newcommand{\xicppi}{ \Xi_c^{\prime0}\pi^{+} }
\newcommand{\xicpipi}{ \Xi_c^{+}\pi^{-}\pi^{+} }
\newcommand{\xicgpi}{ \Xi_c^{0}\gamma\pi^{+} }
\newcommand{\Xicspi}{ \Xi_{c}(2970)^+ \ra \xicspi }
\newcommand{\Xicppi}{ \Xi_{c}(2970)^+ \ra \xicppi }
\newcommand{\Xicpipi}{ \Xi_{c}(2970)^+ \ra \Xi_c(2645)^{0}\pi^{+} \ra \Xi_c^{+}\pi^{-}\pi^{+}}
\newcommand{\Xicgpi}{ \Xi_{c}(2970)^+ \ra \xicppi \ra \xicgpi}

\newcommand{\risodet}{1.67 \pm 0.29\mathrm{(stat.)}^{ +0.15}_{ -0.09}\mathrm{(syst.)} \pm 0.25\mathrm{(IS)}}
\newcommand{\rbl}{{\mathcal{B}[ \Xicspi ]} / { \mathcal{B}[ \Xicppi ] }}

\begin{document}

\preprint{\vbox{ \hbox{   }
			\hbox{Belle Preprint 2020-09}
			\hbox{KEK Preprint 2020-09}
}}

\title{ First Determination of the Spin and Parity of a Charmed-Strange Baryon, ${\bf \Xic}$ }
\noaffiliation
\affiliation{University of the Basque Country UPV/EHU, 48080 Bilbao}
\affiliation{University of Bonn, 53115 Bonn}
\affiliation{Brookhaven National Laboratory, Upton, New York 11973}
\affiliation{Budker Institute of Nuclear Physics SB RAS, Novosibirsk 630090}
\affiliation{Faculty of Mathematics and Physics, Charles University, 121 16 Prague}
\affiliation{Chonnam National University, Gwangju 61186}
\affiliation{University of Cincinnati, Cincinnati, Ohio 45221}
\affiliation{Deutsches Elektronen--Synchrotron, 22607 Hamburg}
\affiliation{Duke University, Durham, North Carolina 27708}
\affiliation{University of Florida, Gainesville, Florida 32611}
\affiliation{Department of Physics, Fu Jen Catholic University, Taipei 24205}
\affiliation{Key Laboratory of Nuclear Physics and Ion-beam Application (MOE) and Institute of Modern Physics, Fudan University, Shanghai 200443}
\affiliation{Gifu University, Gifu 501-1193}
\affiliation{II. Physikalisches Institut, Georg-August-Universit\"at G\"ottingen, 37073 G\"ottingen}
\affiliation{SOKENDAI (The Graduate University for Advanced Studies), Hayama 240-0193}
\affiliation{Gyeongsang National University, Jinju 52828}
\affiliation{Department of Physics and Institute of Natural Sciences, Hanyang University, Seoul 04763}
\affiliation{University of Hawaii, Honolulu, Hawaii 96822}
\affiliation{High Energy Accelerator Research Organization (KEK), Tsukuba 305-0801}
\affiliation{J-PARC Branch, KEK Theory Center, High Energy Accelerator Research Organization (KEK), Tsukuba 305-0801}
\affiliation{Higher School of Economics (HSE), Moscow 101000}
\affiliation{Forschungszentrum J\"{u}lich, 52425 J\"{u}lich}
\affiliation{Hiroshima Institute of Technology, Hiroshima 731-5193}
\affiliation{IKERBASQUE, Basque Foundation for Science, 48013 Bilbao}
\affiliation{Indian Institute of Science Education and Research Mohali, SAS Nagar, 140306}
\affiliation{Indian Institute of Technology Bhubaneswar, Satya Nagar 751007}
\affiliation{Indian Institute of Technology Guwahati, Assam 781039}
\affiliation{Indian Institute of Technology Hyderabad, Telangana 502285}
\affiliation{Indian Institute of Technology Madras, Chennai 600036}
\affiliation{Indiana University, Bloomington, Indiana 47408}
\affiliation{Institute of High Energy Physics, Chinese Academy of Sciences, Beijing 100049}
\affiliation{Institute of High Energy Physics, Vienna 1050}
\affiliation{Institute for High Energy Physics, Protvino 142281}
\affiliation{INFN - Sezione di Napoli, 80126 Napoli}
\affiliation{INFN - Sezione di Torino, 10125 Torino}
\affiliation{Advanced Science Research Center, Japan Atomic Energy Agency, Naka 319-1195}
\affiliation{J. Stefan Institute, 1000 Ljubljana}
\affiliation{Institut f\"ur Experimentelle Teilchenphysik, Karlsruher Institut f\"ur Technologie, 76131 Karlsruhe}
\affiliation{Department of Physics, Faculty of Science, King Abdulaziz University, Jeddah 21589}
\affiliation{Kitasato University, Sagamihara 252-0373}
\affiliation{Korea Institute of Science and Technology Information, Daejeon 34141}
\affiliation{Korea University, Seoul 02841}
\affiliation{Kyoto Sangyo University, Kyoto 603-8555}
\affiliation{Kyungpook National University, Daegu 41566}
\affiliation{Universit\'{e} Paris-Saclay, CNRS/IN2P3, IJCLab, 91405 Orsay}
\affiliation{P.N. Lebedev Physical Institute of the Russian Academy of Sciences, Moscow 119991}
\affiliation{Faculty of Mathematics and Physics, University of Ljubljana, 1000 Ljubljana}
\affiliation{Ludwig Maximilians University, 80539 Munich}
\affiliation{Luther College, Decorah, Iowa 52101}
\affiliation{University of Maribor, 2000 Maribor}
\affiliation{Max-Planck-Institut f\"ur Physik, 80805 M\"unchen}
\affiliation{School of Physics, University of Melbourne, Victoria 3010}
\affiliation{University of Mississippi, University, Mississippi 38677}
\affiliation{University of Miyazaki, Miyazaki 889-2192}
\affiliation{Moscow Physical Engineering Institute, Moscow 115409}
\affiliation{Graduate School of Science, Nagoya University, Nagoya 464-8602}
\affiliation{Universit\`{a} di Napoli Federico II, 80126 Napoli}
\affiliation{Nara Women's University, Nara 630-8506}
\affiliation{National Central University, Chung-li 32054}
\affiliation{National United University, Miao Li 36003}
\affiliation{Department of Physics, National Taiwan University, Taipei 10617}
\affiliation{H. Niewodniczanski Institute of Nuclear Physics, Krakow 31-342}
\affiliation{Nippon Dental University, Niigata 951-8580}
\affiliation{Niigata University, Niigata 950-2181}
\affiliation{Novosibirsk State University, Novosibirsk 630090}
\affiliation{Osaka City University, Osaka 558-8585}
\affiliation{Pacific Northwest National Laboratory, Richland, Washington 99352}
\affiliation{Panjab University, Chandigarh 160014}
\affiliation{Peking University, Beijing 100871}
\affiliation{University of Pittsburgh, Pittsburgh, Pennsylvania 15260}
\affiliation{Punjab Agricultural University, Ludhiana 141004}
\affiliation{Research Center for Nuclear Physics, Osaka University, Osaka 567-0047}
\affiliation{RIKEN BNL Research Center, Upton, New York 11973}
\affiliation{Department of Modern Physics and State Key Laboratory of Particle Detection and Electronics, University of Science and Technology of China, Hefei 230026}
\affiliation{Seoul National University, Seoul 08826}
\affiliation{Showa Pharmaceutical University, Tokyo 194-8543}
\affiliation{Soochow University, Suzhou 215006}
\affiliation{Soongsil University, Seoul 06978}
\affiliation{Sungkyunkwan University, Suwon 16419}
\affiliation{School of Physics, University of Sydney, New South Wales 2006}
\affiliation{Department of Physics, Faculty of Science, University of Tabuk, Tabuk 71451}
\affiliation{Tata Institute of Fundamental Research, Mumbai 400005}
\affiliation{Department of Physics, Technische Universit\"at M\"unchen, 85748 Garching}
\affiliation{School of Physics and Astronomy, Tel Aviv University, Tel Aviv 69978}
\affiliation{Toho University, Funabashi 274-8510}
\affiliation{Earthquake Research Institute, University of Tokyo, Tokyo 113-0032}
\affiliation{Department of Physics, University of Tokyo, Tokyo 113-0033}
\affiliation{Tokyo Institute of Technology, Tokyo 152-8550}
\affiliation{Tokyo Metropolitan University, Tokyo 192-0397}
\affiliation{Utkal University, Bhubaneswar 751004}
\affiliation{Virginia Polytechnic Institute and State University, Blacksburg, Virginia 24061}
\affiliation{Wayne State University, Detroit, Michigan 48202}
\affiliation{Yamagata University, Yamagata 990-8560}
\affiliation{Yonsei University, Seoul 03722}
  \author{T.~J.~Moon}\affiliation{Seoul National University, Seoul 08826} 
  \author{K.~Tanida}\affiliation{Advanced Science Research Center, Japan Atomic Energy Agency, Naka 319-1195} 
  \author{Y.~Kato}\affiliation{Graduate School of Science, Nagoya University, Nagoya 464-8602} 
  \author{S.~K.~Kim}\affiliation{Seoul National University, Seoul 08826} 
  \author{I.~Adachi}\affiliation{High Energy Accelerator Research Organization (KEK), Tsukuba 305-0801}\affiliation{SOKENDAI (The Graduate University for Advanced Studies), Hayama 240-0193} 
  \author{J.~K.~Ahn}\affiliation{Korea University, Seoul 02841} 
  \author{H.~Aihara}\affiliation{Department of Physics, University of Tokyo, Tokyo 113-0033} 
  \author{S.~Al~Said}\affiliation{Department of Physics, Faculty of Science, University of Tabuk, Tabuk 71451}\affiliation{Department of Physics, Faculty of Science, King Abdulaziz University, Jeddah 21589} 
  \author{D.~M.~Asner}\affiliation{Brookhaven National Laboratory, Upton, New York 11973} 
  \author{V.~Aulchenko}\affiliation{Budker Institute of Nuclear Physics SB RAS, Novosibirsk 630090}\affiliation{Novosibirsk State University, Novosibirsk 630090} 
% ...
\collaboration{The Belle Collaboration}

\begin{abstract}
We report results from a study of the spin and parity 
of $\Xic$ using a $980~\mathrm{fb^{-1}}$ data sample 
collected by the Belle detector at the KEKB asymmetric-energy $e^{+}e^{-}$ collider.
The decay angle distributions in the chain $\Xicpipi$ are analyzed to determine the spin of this charmed-strange baryon.
The angular distributions strongly favor the $\Xic$ spin 
$J =1/2$ over $3/2$ or $5/2$, 
under an assumption that the lowest partial wave dominates in the decay.
We also measure the ratio of $\Xic$ decay branching fractions 
$R=\rbl=\risodet$, 
where the last uncertainty is due to possible isospin-symmetry-breaking effects. 
This $R$ value favors the spin-parity $J^P=1/2^+$ with the spin of the light-quark degrees of freedom $s_{l}=0$.
This is the first determination of the spin and parity of a charmed-strange baryon.
\end{abstract}
\maketitle

\tighten

{\renewcommand{\thefootnote}{\fnsymbol{footnote}}}
\setcounter{footnote}{0}

Charmed-strange baryons comprise one light (up or down) quark, 
one strange quark, and a more massive charm quark. 
They provide an excellent laboratory to test various theoretical 
models, in which the three constituent quarks are effectively described in terms of 
a heavy quark plus a light diquark system \cite{ebert, chen1}.
The ground and excited states of $\xic$ baryons have 
been observed during the last few decades \cite{Kato:2018ijx}.
At present there is no experimental determination of their spins or parities.

Excited $\xic$ states with an excitation energy less than $400\mev$
can be uniquely identified as particular states predicted by the quark model \cite{pdg}.
However, in the higher excitation region,  
there are multiple states within the typical mass accuracy of 
quark-model predictions of around $50\mev$, making a unique identification challenging.
In order to identify and understand the nature of excited 
$\xic$ baryons, experimental determination of their spin-parity is indispensable.

In this Letter, we report the first measurement of the spin-parity of a $\xic$ baryon.
We choose $\Xi_c(2970)$, earlier known as $\Xi_c(2980)$, 
an excited state of the lightest charmed-strange baryons, 
for which a plausible spin-parity assignment is not given by the Particle Data Group~\cite{pdg}. 
It was first observed in the decay mode $\Lambda_c^+ \bar{K} \pi$ by Belle~\cite{chistov} 
and later confirmed by BaBar~\cite{aubert} in the same decay mode. 
It was also observed in the $\xic(2645) \pi$ channel at Belle~\cite{lesiak}.
Its mass and width have been precisely measured with a larger data sample 
using the $\xic(2645)\pi$ channel by a recent study~\cite{yelton}, 
which also observed the decay mode $\xic^\prime\pi$ for the first time.
The high statistics of the Belle data, especially for the $\xic(2645) \pi$ channel, 
recorded in a clean $e^{+}e^{-}$ environment provides 
an ideal setting for the experimental determination 
of the spin and parity of charmed-strange baryons.

Theoretically, there are many possibilities for the spin-parity assignment of $\xic(2970)$.
For example, a quark-model calculation by Roberts and Pervin \cite{roberts}
listed $J^P=1/2^+$, $3/2^+$, $5/2^+$, and $5/2^-$ as possible candidates.
Similarly, most quark-model-based calculations predict the $\Xi_c(2970)$ as a
$2S$ state with $J^P=1/2^+$ or $3/2^+$~\cite{ebert, garcilazo, chen1,
shah, gandhi}, while some of them find negative parity states in the close vicinity \cite{ebert,migura}.
 There are even calculations that directly assign negative 
parity to the $\Xi_c(2970)$ \cite{nieves, chen2}. 
The unclear theoretical situation motivates
an experimental determination
of the spin-parity of the $\Xic$ that will provide important information to test these predictions 
and help decipher the nature of the state.

In this study, the spin is determined by testing 
possible spin hypotheses of $\Xic$ with angular analysis of the decay $\Xicpipi$. 
Similarly, its parity is established from the 
ratio of branching fractions of the two decays, $\Xicspi$ and $\Xicppi$. 
We note that recently LHCb observed two new states in the $\Lambda_c^+ K^-$ channel~\cite{lhcb} and a narrow third state 
$\Xi_c(2965)$, which is very close in mass to the much wider $\Xi_c(2970)$.
It is however assumed, because of their significantly different widths and different decay channels 
in which they are observed, that they are two different states.
In this work, it is assumed that the peak structures observed in $\Xi_c(2645) \pi$ and $\Xi_c^\prime \pi$ channels come from a single resonance.

The analysis is based on a sample of $e^+e^-$ annihilation data totaling an integrated luminosity of $980~\mathrm{fb^{-1}}$ 
recorded by the Belle detector \cite{Belle} at the KEKB asymmetric-energy $e^+e^-$ collider \cite{KEKB}. 
Belle was a large-solid-angle magnetic
spectrometer consisting of a silicon vertex detector (SVD),
a 50-layer central drift chamber (CDC), an array of
aerogel threshold Cherenkov counters,  
a barrel-like arrangement of time-of-flight
scintillation counters, and an electromagnetic calorimeter
comprised CsI(Tl) crystals, all located inside 
a superconducting solenoid coil that provided a 1.5~T
magnetic field.
An iron flux return placed outside of
the coil was instrumented to detect $K_L^0$ mesons and muons.
Two inner-detector configurations were used: a 2.0 cm radius beampipe
and a three-layer SVD were used for the first sample
of $156~\mathrm{fb^{-1}}$, while a 1.5 cm radius beampipe, a four-layer
SVD and a small-cell inner CDC were used to record  
the remaining $824~\mathrm{fb^{-1}}$\cite{svd2}. 
Using a GEANT-based Monte Carlo (MC) simulation \cite{geant3},
the detector response and its acceptance are modeled to study 
the mass resolution of signals and obtain reconstruction efficiencies.

The $\Xic$ is reconstructed in the two decay modes, 
$\xicspi$ and $\Xi_{c}^{'0} \pi^{+}$ 
with $\xics \to \xic^+\pi^-$ and $\xicp \to \xic^0 \gamma$,
closely following the earlier analysis by Belle~\cite{yelton}.
The only difference is that $\Xi_{c}^{+}$ and $\Xi_{c}^{0}$ are 
reconstructed in the decay modes $\xic^+ \ra \Xi^{-} \pi^{+} \pi^{+}$ and
$\Xi_{c}^{0} \ra \Xi^{-} \pi^{+}/\Omega^{-} K^{+}$ 
[with $\Xi^- (\Omega^-) \to \Lambda \pi^- (K^-)$ and $\Lambda \to p\pi^-$],
which have high statistics with good signal-to-background ratios.
The scaled momentum $x_{p}$ = $p^{\ast}c$/$\sqrt{s/4-m^{2}c^2}$, where
$p^{\ast}$ is the center-of-mass (c.m.) momentum of the $\Xic$ candidate, $\sqrt{s}$ is the total c.m. energy, 
and $m$ is the mass of the $\Xic$ candidate, is required to be greater than 0.7.

%



In order to determine the spin of $\Xic$, 
two angular distributions of the decay chain
$\Xic\ra\xics\pi_{1}^{+}\ra\xic^{+}\pi_{2}^{-}\pi_{1}^{+}$ are analyzed.
The first one is the helicity angle $\theta_h$ of $\Xic$, 
defined as the angle between the direction of the primary pion $\pi_1^{+}$
and the opposite of boost direction of the c.m. frame, both calculated in the rest frame of the $\Xic$. 
Such an angle was used to determine the spin of $\Lambda_c(2880)^+$ \cite{mizuk}.
The second one is the helicity angle of $\xics$, 
defined as the angle between the direction of the secondary pion $\pi_2^{-}$
and the opposite direction of the $\Xic$, both calculated in the rest frame of the $\xics$. 
This angle, referred to as $\theta_c$, represents
angular correlations of the two pions, 
because $\pi_1^{+}$ and $\xics$ are emitted back to back in the rest frame of $\Xic$.

The angular distributions are obtained by dividing the data into 10 equal bins for $\cos\theta_h$ and $\cos\theta_c$, each extending for intervals of 0.2.
For each $\cos\theta_h$ or $\cos\theta_c$ bin, 
the yield of $\Xicspi$ is obtained by fitting 
the invariant-mass distribution of $M(\xicpipi)$ 
for the $\xics$ signal region and sidebands. 
These two regions are defined as 
$\vert M(\xic^+\pi^-)-m[\xics]\vert < 5\mevcc$ and
$15\mevcc < \vert M(\xic^+\pi^-)-m[\xics]\vert < 25\mevcc$,
respectively, with $m[\xics]=2646.38 \mevcc$ \cite{pdg}.
To consider the nonresonant contribution, which is the direct three-body decay into $\xicpipi$, 
a sideband subtraction is performed.
The $\Xic$ signal is parametrized by a Breit-Wigner function
convolved with a double-Gaussian resolution function 
and the background by a first-order polynomial.
Parameters for the Breit-Wigner are fixed to 
the values from the previous Belle measurement \cite{yelton} 
while those for the resolution function are determined from an MC simulation. 
The yields obtained from the fits and efficiencies determined from signal MC events are given in Ref.~\cite{supple}.
\par
The following systematic uncertainties are considered for 
each $\cos\theta_h$ and $\cos\theta_c$ bin.
The resultant systematic uncertainties in the yield of each bin are presented in parentheses.
The uncertainty due to the resolution function is 
checked by changing the 
width of the core Gaussian component by 10$\%$ to consider a possible 
data-MC difference in resolution ($0.2\%$ at most). 
Also, each resolution parameter is varied within its 
statistical uncertainty determined from signal MC events ($0.1\%$ at most). 
The statistical uncertainty in the efficiency is negligible.
The uncertainty due to the background model is determined by 
redoing the fit with a second-order polynomial or constant function
instead of the first-order polynomial ($0.7$ -- $47\%$).
The uncertainty coming from the mass and width of $\Xic$ is determined by 
changing their values within uncertainties~\cite{yelton} ($6.7$ -- $12\%$).
All of these uncertainties are added in quadrature ($6.7$ -- $47\%$).

\begin{figure}[t]
	\includegraphics[height=0.35\textwidth]{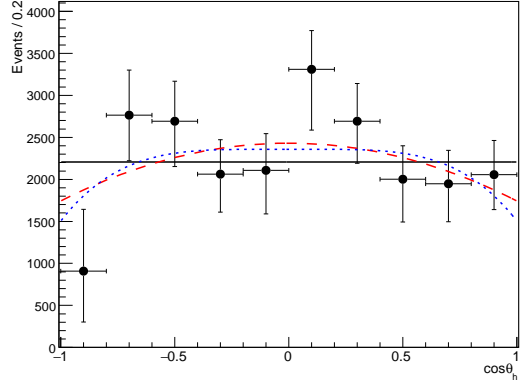}
\vspace*{-5mm}
\caption{Yields of the $\Xicspi$ decay as a function of 
$\cos\theta_h$ after the sideband subtraction and efficiency correction. 
Points with error bars are data that include the quadrature sum
 of statistical and systematic uncertainties. 
The fit results with $W_{1/2}$ (solid black), 
$W_{3/2}$ (dashed red), and $W_{5/2}$ (dotted blue) are overlaid.}
\label{fig:angle_cosh}
\end{figure}

Yields of the decay $\Xicspi$ after the $\xics$ sideband subtraction and efficiency 
correction are shown as a function of $\cos\theta_h$ in Fig. \ref{fig:angle_cosh}. 
Although the quantum numbers of the $\Xi_c(2645)$ 
have not yet been measured, in the quark model the natural assumption 
for its spin-parity is $J^P={3/2}^+$.
Then the expected decay-angle distributions $W_J$
for spin hypotheses of $J=1/2$, $3/2$, and $5/2$ for $\Xic$
are as follows \cite{pilkuhn}:
\begin{eqnarray}
W_{1/2} &=& \rho_{11} = \frac{1}{2}\\
\nonumber
W_{3/2} &=& \rho_{33}\left\{1+T\left(\frac{3}{2}\cos^2\theta_h-\frac{1}{2}\right)\right\} \\ 
&& + \rho_{11}\left\{1+T\left(-\frac{3}{2}\cos^2\theta_h+\frac{1}{2}\right)\right\} \\
\nonumber
W_{5/2} &=& \frac{3}{32}[\rho_{55}5\{(-\cos^4\theta_h-2\cos^2\theta_h+3) \\ \nonumber
&& \hspace{5mm}+T(-5\cos^4\theta_h+6\cos^2\theta_h-1)\} \\ \nonumber
&& \hspace{0mm}+\rho_{33}\{(15\cos^4\theta_h-10\cos^2\theta_h+11) \\ \nonumber
&& \hspace{5mm}+T(75\cos^4\theta_h-66\cos^2\theta_h+7) )\} \\ \nonumber
&& \hspace{0mm}+\rho_{11}2\{(-5\cos^4\theta_h+10\cos^2\theta_h+3)  \\ 
&& \hspace{5mm}+T(-25\cos^4\theta_h+18\cos^2\theta_h-1)\}] .
\end{eqnarray}
Here, $T = \frac{\vert{{\cal T}(p,\frac{3}{2},0)}\vert^2-\vert{{\cal T}(p,\frac{1}{2},0)}\vert^2}
{\vert{{\cal T}(p,\frac{3}{2},0)}\vert^2+\vert{{\cal T}(p,\frac{1}{2},0)}\vert^2}$ 
and ${\cal T}(p,\lambda_1,\lambda_2)$ is the matrix element of a two-body decay 
with the momentum $p$ of the daughters in the mother's rest frame and the helicities of daughters
being $\lambda_1$ for $\xics$ and $\lambda_2$ for $\pi^+$. 
The parameter $\rho_{ii}$ is the diagonal element of the spin-density matrix of $\Xic$ with helicity $i/2$.
The sum of $\rho_{ii}$ for positive odd integer $i$ is normalized to $1/2$.

\begin{table}[t] 
  \begin{center} 
  \caption{Result of the angular analysis of the decay $\Xicspi$. 
 Here, n.d.f. denotes the number of degrees of freedom.}
  \label{tbl:fit_cosh} 
 \begin{tabular}{  c  | c  c   c  }
    \hline\hline
	 Spin hypothesis & ~~~1/2~~~ & ~~~3/2~~~ & ~~~5/2~~~ \\
	\hline
	$\chi^2/\mathrm{n.d.f.}$	& 9.3/9		& 7.7/7			& 7.5/6 \\ 
	Probability			& 41\%		& 36\%			& 28\%	\\ 
	T				& --		& $-0.5 \pm 1.1$	& $0.7 \pm 1.6$  \\
	$\rho_{11}$			& 0.5		& ~$0.13 \pm 0.26$~	& ~$0.08 \pm 0.27$~  \\
	$\rho_{33}$			& --		& $0.37 \pm 0.26$	& $0.12 \pm 0.09$ \\
	$\rho_{55}$			& --		& --			& $0.30 \pm 0.28$ \\
	\hline\hline
    \end{tabular}
\end{center}
\end{table}

The fit results are summarized in Table \ref{tbl:fit_cosh}.
Though the best fit is obtained for the spin 1/2 hypothesis,  
the exclusion level of the spin 3/2 (5/2) hypothesis is 
as small as 0.8 (0.5) standard deviations. 
Therefore, the result is inconclusive.
In other words, it is consistent
with a uniform distribution, which can be exhibited by any
spin $J$ if the initial state is unpolarized.

\begin{figure}[htb]
	\includegraphics[height=0.35\textwidth]{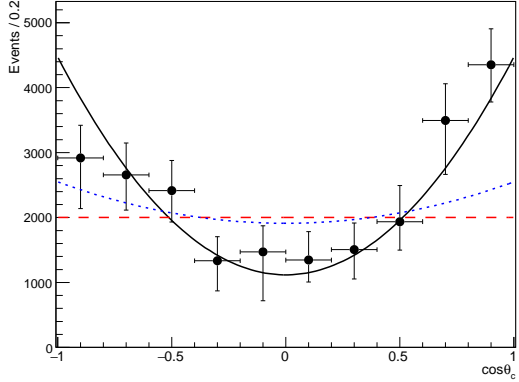}
\vspace*{-5mm}	
\caption{The yields of $\Xicpipi$ decay as a function of $\cos\theta_c$. 
The fit results with spin-parity hypotheses 
$\frac{1}{2}^\pm$ (solid black), $\frac{3}{2}^-$ (dashed red), and 
$\frac{5}{2}^+$ (dotted blue) are also presented.}
\label{fig:angle_cosc_pw}
\end{figure}

In order to draw a more decisive conclusion, 
we further analyze the angular correlations 
of the two pions in the $\Xicpipi$ decay. 
In this case, the expected angular distribution is \cite{pilkuhn}
\begin{equation}
		W(\theta_c) = \frac{3}{2}\left[\rho^*_{33}\sin^2\theta_c + 
\rho^*_{11}\left(\frac{1}{3}+\cos^2\theta_c\right)\right], \label{eq:ExpectedCorrelation}
\end{equation}
where $\rho^*_{ii}$ is the diagonal element of the spin-density matrix of $\xics$
with the normalization condition $\rho^*_{11}+\rho^*_{33}=1/2$.
Figure \ref{fig:angle_cosc_pw} shows the yields of $\Xic$ as a
function of $\cos\theta_c$ after the $\xics$ sideband subtraction and efficiency correction. 
A fit to the expected distribution [Eq. (\ref{eq:ExpectedCorrelation})] 
gives  a good $\chi^2/{\rm n.d.f.}=5.6/8$ with 
$\rho^*_{11}=0.46\pm0.04$ and $\rho^*_{33}=0.5-\rho^*_{11}=0.04\pm0.04$,
which indicates that the population of helicity 3/2 state 
is consistent with zero. 
This result is most consistent with
the spin 1/2 hypothesis of $\Xic$, as only the helicity 1/2 state of $\xics$ 
can survive due to helicity conservation. 
Indeed, assuming that the lowest partial wave dominates for the 
$\Xicspi$ decay, the expected angular correlations can be calculated 
as summarized in Table \ref{tbl:angle_dist} \cite{Jafar}. 
Fitting the data 
to the cases $J^P=1/2^\pm$, $3/2^-$, and $5/2^+$, we obtain the fit results as summarized in Table~\ref{tbl:fit_cosc_supp}.
We find the result to favor the $1/2^{\pm}$ hypothesis over the  $3/2^-$ ($5/2^+$) one 
at the level of 5.1 (4.0) standard deviations.
The exclusion level is even higher for the other hypotheses
for which the expected angular distributions are upwardly convex.
We note that this result also excludes the $\xic(2645)$ spin of $1/2$ in which the distribution should be flat,
and that the present discussion is still true even if there are two resonances, $\Xi_c(2970)$ and $\Xi_c(2965)$ ~\cite{lhcb}.

\begin{table}[htb] 
  \begin{center} 
  \caption{ Expected angular distribution for spin-parity 
hypotheses of $\Xic$ with an assumption that the lowest partial wave dominates.}
  \label{tbl:angle_dist} 
\begin{tabular}  {@{\hspace{0.2cm}}l@{\hspace{0.2cm}}  @{\hspace{0cm}}c@{\hspace{0cm}} @{\hspace{0cm}}c@{\hspace{0cm}}}
    \hline\hline
	~$J^P$~ & ~Partial Wave~ & ~~$W(\theta_c)$~~  \\
	\hline
	$1/2^+$	&	$P$ 		& $1+3\cos^2\theta_c$ \\ 
	$1/2^-$	&	$D$		& $1+3\cos^2\theta_c$ \\ 
	$3/2^+$	&	$P$		& $1+6\sin^2\theta_c$ \\
	$3/2^-$	&	$S$		& $1$		            \\
	$5/2^+$	&	$P$		& $1+(1/3)\cos^2\theta_c$	 \\
	$5/2^-$	&	$D$		& $1+(15/4)\sin^2\theta_c$ \\
	\hline\hline
    \end{tabular}
\end{center}
\end{table}

\begin{table}[!htb] 
  \begin{center} 
  \caption{Results of the angular analysis of the decay $\Xicspi$ 
with an assumption that the lowest partial wave dominates. }
  \label{tbl:fit_cosc_supp} 
 \begin{tabular}
 {@{\hspace{0.3cm}}l@{\hspace{0.3cm}} | @{\hspace{0.2cm}}c@{\hspace{0.2cm}} @{\hspace{0.2cm}}c@{\hspace{0.2cm}} @{\hspace{0.2cm}}c@{\hspace{0.2cm}} }

    \hline\hline
	~$J^P$~ & ~~~$1/2^\pm$~~~ & ~~~$3/2^{-}$~~~ & ~~~$5/2^{+}$~~~ \\
	\hline
	$\chi^2/\mathrm{n.d.f.}$	& 6.4/9	& 32.2/9		& 22.3/9 \\ 
	Probablility			& 0.69	& 1.8$\times10^{-4}$	& $7.9\times10^{-3}$	\\
	\hline\hline
    \end{tabular}
\end{center}
\end{table}

The ratio of branching fractions 
$R=    {\cal B}[\Xi_{c}(2970)^{+} \to  \xics\pi^+]/{\cal B}[ \Xi_{c}(2970)^{+} \to \xicp\pi^+]$ 
is sensitive to the parity of $\Xic$ \cite{cheng,mizuk}. 
In principle, the $R$ value can be determined using the following equation: 
\begin{equation}
	 R=\left.{ \frac{N^{\ast}}{\mathcal{E}^{\ast}\times {\cal B}^+ }  } 
             \right/{\frac{N^{\prime}}{\sum_{i}\mathcal{E}^{\prime}_{i} \times {\cal B}^{0}_i }  } ,
\label{eq:rdef}
\end{equation}
where $N^\ast$($N^\prime$) is the yield of $\Xic$ in the $\xicspi$ ($\xicppi$) decay mode,
$\mathcal{E} ^\ast$ ($\mathcal{E} ^\prime_i$)  is the reconstruction efficiency 
of $\Xic$ for the decay $\xicspi$ ($\xicppi$ with $i=\Xi^-\pi^+$ or $\Omega^-K^+$ mode of $\Xi_{c}^{0}$), and
${\cal B}^{+}$ (${\cal B}_{i}^{0}$) is the measured branching fraction of 
$\Xi_{c}^{+} \to \Xi^{-} \pi^{+} \pi^{+}$ ($\Xi_{c}^{0} \to i$-th subdecay mode)
\cite{ybli_p, ybli_0, xic0_br}.
In this case, however, the uncertainty will be dominated by the 
branching fractions of the ground-state $\xic$ baryons. 
Such uncertainties are avoided by calculating the ratio in a different way, with inclusive measurements of $\xic^+$ and $\xic^0$ and 
an assumption of isospin symmetry in their inclusive cross sections.
We note that this assumption is confirmed 
within 15\% in the $\Sigma_c^{(*)}$ case \cite{Belle-Sc}.

\begin{figure}[t]
	\includegraphics[height=0.35\textwidth]{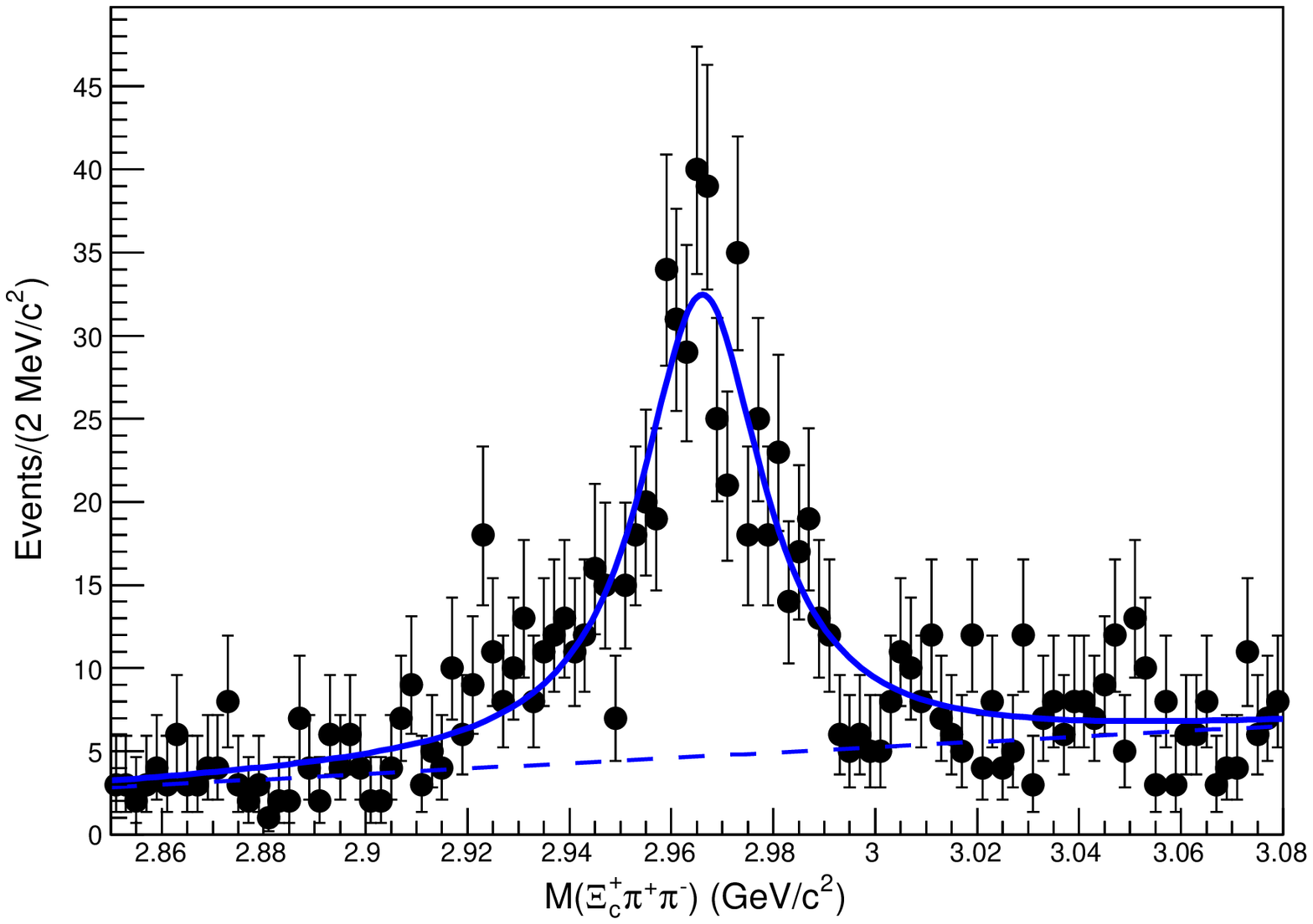}
\vspace*{-5mm}
\caption{ $\xicpipi$ invariant-mass distribution for the decay $\Xicpipi$. 
Black points with error bars are data. 
The fit result (solid blue curve) is also presented along with 
the background (dashed blue curve).}
\label{fig:mass_xic2970}
\end{figure}

\begin{figure}[t]
	\includegraphics[height=0.35\textwidth]{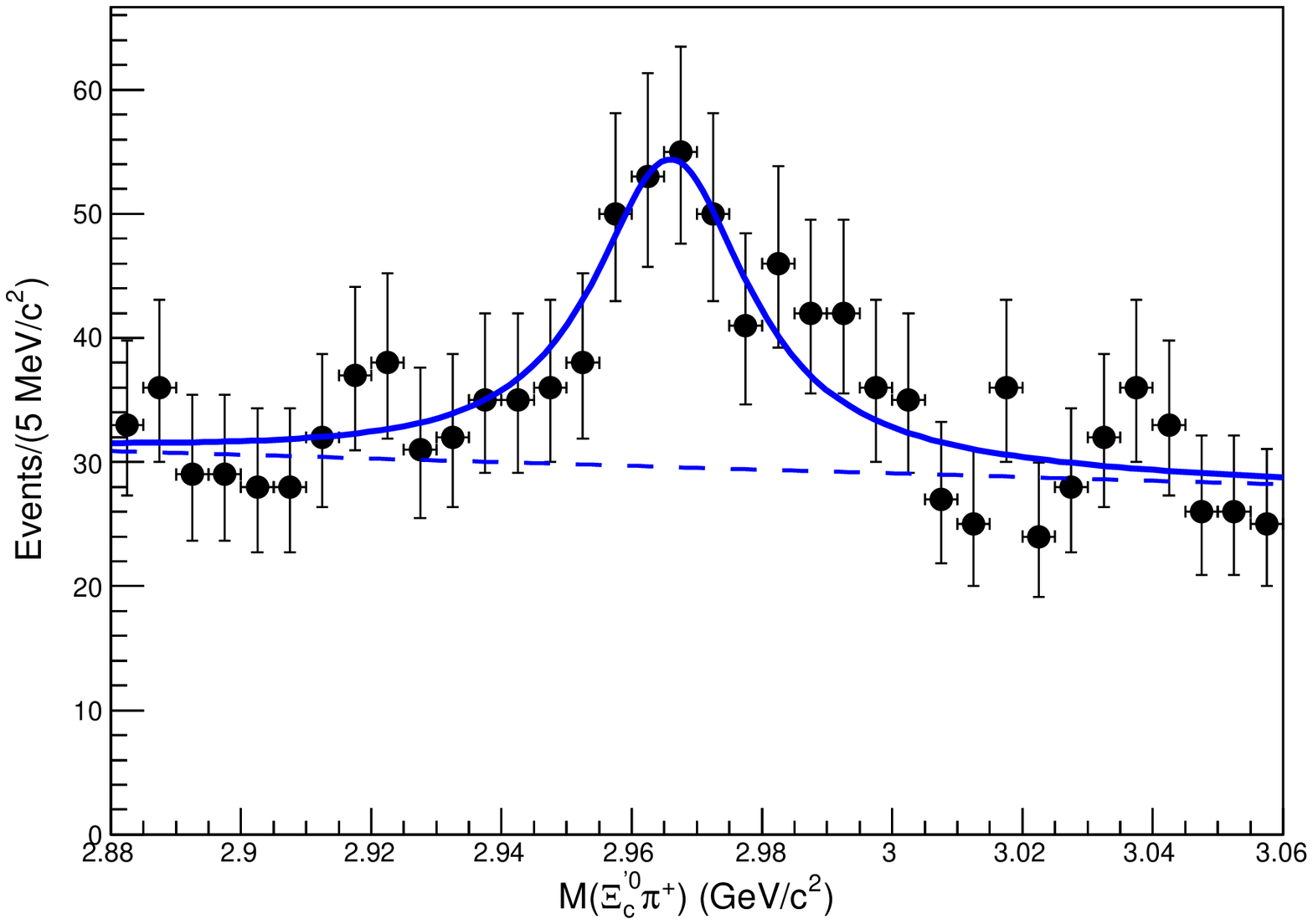}
\vspace*{-5mm}
\caption{$\xicppi$ invariant-mass distribution for the decay $\Xicgpi$.
Black points with error bars are data.
The fit result (solid blue curve) is also presented along with 
the background (dashed blue curve).}
\label{fig:mass_xic2970prime}
\end{figure}

The branching fraction of $\Xi_{c}^{+(0)}$ in a certain subdecay mode is given as
\begin{equation}
{\cal B}_{i}^{+(0)} = \frac{N(\Xi_{c}^{+(0)})_i}{\mathcal{L} \times \sigma_{\xic} \times \epsilon^{+(0)}_{i}},
\label{eq:replace}
\end{equation}
where $N({\Xi_c}^{+(0)})_{i}$ and $\epsilon_{i}^{+(0)}$ are 
the yield and reconstruction efficiency of the 
$\Xi_{c}^{+(0)}$ ground states for the $i$-th subdecay mode,
$\mathcal{L}$ is the integrated luminosity, 
and $\sigma_{\xic}$ is the inclusive production cross section of $\xic$ 
which is assumed to be the same for $\Xi_{c}^{0}$ and $\Xi_{c}^{+}$.
By replacing the ground-state $\Xi_{c}$ branching fractions in Eq. (\ref{eq:rdef})
with the values in Eq. (\ref{eq:replace}), $R$ can be rewritten as
\begin{equation}
        	    R=\left.{ \frac{N^{\ast}}{\mathcal{E}^{\ast}\times \frac{N(\xic^+)}{\epsilon^{+}} }  } \right/{\frac{N^{\prime}}{\sum_i \mathcal{E}^{\prime}_{i} \times \frac{N(\xic^0)_i}{\epsilon^{0}_i} }  }.
\end{equation}
Here, $N^{\ast}$ and $N^{\prime}$ are obtained by fitting the $\xicspi$ 
and $\xicppi$ invariant-mass distributions
with all the phase space integrated.
For the $\xicspi$ channel, a sideband subtraction is performed.
For the $\xicppi$ channel, the fit is performed for the $\xicp$ signal region, defined as 
$\vert M(\xic^0\gamma)-m[\xicp]\vert < 8\mevcc$ with $m[\xicp]=2579.2 \mevcc$ \cite{pdg}.
For both decay channels, we perform fits using
a Breit-Wigner function convolved with a double Gaussian as signal and a first-order polynomial as background.
The invariant-mass distributions together with the fit results 
are shown in  Figs. \ref{fig:mass_xic2970} and \ref{fig:mass_xic2970prime}.
Similarly, $N(\Xi_{c}^{+/0})$ are obtained by fitting the invariant-mass distributions of $\xic$ candidates.
Ground-state $\xic$ baryons are reconstructed in a similar way as $\Xic$;
the only difference being 
that $x_{p}$ is calculated with the mass of $\xic$ and required to be greater than 0.6.
The fit is performed with a double-Gaussian function as signal and a first-order polynomial as background.

The following systematic uncertainties are considered for the $R$ measurement.
The uncertainty coming from the resolution function is checked by changing the 
width of the core Gaussian component by 10$\%$ to consider possible data-MC difference in resolution (${+3.3\%}/{-3.4\%}$). 
Also, each parameter is varied within its statistical uncertainty determined from signal MC events (0.4$\%$). 
The statistical uncertainty in the efficiency is negligible. 
The mass and width of $\Xic$ are changed within their uncertainties~\cite{yelton} (${+4.1\%}/{-1.7\%}$). 
The uncertainty due to the background shape is determined by changing it from 
a first-order polynomial to a constant function and second-order polynomial (${+6.8\%}/{-0.9\%}$). 
The uncertainty due to the tracking efficiency is 0.35$\%$ per track.
The systematic uncertainty due to the pion-identification efficiency ($1.2\%$) is obtained 
using $D^{\ast +} \to D^{0} \pi^{+}$ and $D^{0} \to K^{-} \pi^{+}$ decays. 
Similarly, the uncertainty due to $\gamma$ reconstruction is obtained from the $\Sigma^0 \ra \Lambda \gamma$ decay 
and determined to be $3.2\%$. All of these uncertainties are added in quadrature (${+9.2\%}/{-5.2\%}$).

The $R$ value is obtained as  $\risodet$,
where the last uncertainty is due to possible isospin-symmetry-breaking effects ($15\%$).
As a cross check, we have also calculated the same quantity by using the measured branching 
fractions of $\Xi_{c}^{+/0}$ as 
$R=2.05 \pm 0.36({\rm stat.}){^{ +0.18}_{ -0.09}}({\rm syst.}){^{ +1.75}_{ -0.87}}({\rm BF})$, 
where the last uncertainty is due to uncertainties 
in the branching fractions of the ground-state $\xic$ baryons. 
The two values are consistent within uncertainties.
We note that the mass spectra of $\Xic$ in this study can be well described by a single resonance with the mass and width from the previous Belle measurement~\cite{yelton}.


Heavy-quark spin symmetry (HQSS) predicts $R=1.06$ (0.26) for a $1/2^+$ state 
with the spin of the light-quark degrees of freedom $s_l=0$ (1),
as calculated using Eq. (3.17) of Ref. \cite{cheng}.
For the case of $J^P=1/2^-$, we expect $R \ll 1$ because the decay to
$\xicppi$ is in $S$ wave while that to $\xicspi$ is in $D$ wave. 
Therefore, our result favors a positive-parity assignment with $s_l=0$.
We note that HQSS predictions 
could be larger than the quoted value by a factor of $\sim 2$
with higher-order terms in $(1/m_c)$ \cite{hqs_falk_mehen}, so the result is 
consistent with the HQSS prediction for $J^P(s_l)=1/2^+(0)$.

The obtained spin-parity assignment is consistent with 
most quark-model-based calculations \cite{roberts, ebert, gandhi, chen1, shah, gandhi, migura}.
 However, some of them \cite{ebert, gandhi} predict $J^P=1/2^+$ with $s_l=1$
which is inconsistent with our result. 
We note that $J^P=1/2^+$ are the same as those of the Roper resonance $[N(1440)]$~\cite{roper}, $\Lambda(1600)$,
and $\Sigma(1660)$; and interestingly,
their excitation energy levels are the same as that of $\xic(2970)$ ($\sim 500 \mev$) 
even though the quark masses are different. This fact may give a hint 
at the structure of the Roper resonance. 
Therefore, it would be interesting to see if there are further analogous states 
at the same excitation energy in systems with different flavors
such as $\Sigma_c$, $\Lambda_c$, $\Omega_c$, $\Lambda_b$, and $\Xi_b$ baryons.

In summary, we have determined the spin and parity of the $\Xic$ for the first time using 
the decay-angle distributions in $\Xicpipi$ 
and the ratio of $\Xic$ branching fractions of the two decays,
$\Xicdecay$. The decay-angle distributions strongly favor $J=1/2$ assignment over
$3/2$ or $5/2$ under an assumption that
the lowest partial wave dominates in the decay,
and the ratio $R=\risodet$ 
favors $J^P(s_l)=1/2^+(0)$ over the other possibilities.

We thank the KEKB group for excellent operation of the
accelerator; the KEK cryogenics group for efficient solenoid
operations; and the KEK computer group, the NII, and 
PNNL/EMSL for valuable computing and SINET5 network support.  
We acknowledge support from MEXT, JSPS, Nagoya's TLPRC and KAKENHI Grant No. JP19H05148 (Japan);
ARC (Australia); FWF (Austria); NSFC and CCEPP (China); 
MSMT (Czechia); CZF, DFG, EXC153, and VS (Germany);
DST (India); INFN (Italy); 
MOE, MSIP, NRF, RSRI, FLRFAS project, GSDC of KISTI and KREONET/GLORIAD (Korea);
MNiSW and NCN (Poland); MSHE, Agreement 14.W03.31.0026 (Russia); University of Tabuk (Saudi Arabia); ARRS (Slovenia);
IKERBASQUE (Spain); 
SNSF (Switzerland); MOE and MOST (Taiwan); and DOE and NSF (USA).
T. J. Moon and S. K. Kim acknowledge support by NRF Grant No. 2016R1A2B3008343.


\clearpage
\begin{center}
{\large \bf First Determination of the Spin and Parity of a Charmed-Strange Baryon, ${\bf \Xic}$ \\
{\rm\it Supplemental Material}}
\end{center}

For each $\cos\theta_h$ or $\cos\theta_c$ bin, the yield of $\Xicspi$ is obtained by fitting the invariant-mass distribution of $M(\xicpipi)$ for the $\xics$ signal region and sidebands.
for the signal region, the fitted yield are listed in Table \ref{tbl:yield} and 
for the sidebands, the statistics is too small to obtain a reliable yield from fits for the $\cos\theta$ bins.
Total yield from the $\xics$ sidebands is thus averaged over the $\cos\theta$ bins, which gives a yield of $1.0 \pm 0.6$ events for each bin.
For each $\cos\theta_h$ and $\cos\theta_c$ bin, the reconstruction efficiency of $\Xic$ is determined from signal MC events, 
as shown in Table \ref{tbl:eff_angle}.
\begin{table}[!htb]
  \begin{center}
    \caption{ Summary of the yield of $\Xicspi$ obtained by fitting the invariant-mass distribution of  $M(\xicpipi)$ for the $\xics$ signal region for each $\cos\theta_{h}$ and $\cos\theta_{c}$ bin. The uncertainties are statistical.
    }
        \label{tbl:yield}
\renewcommand{\arraystretch}{1.1}
    \begin{tabular}{c | c |  c | c}\hline\hline
      \multirow{2}{*}{$\cos\theta_{h}$~} &  \multirow{2}{*}{~Yield~} &  \multirow{2}{*}{$\cos\theta_{c}$~} & \multirow{2}{*}{~Yield~}\\
      &~& \\
      \hline
      ~~~$-1< \cos\theta_{h}<-0.8$~   & ~$15.6 \pm 9.7$~ &  ~~~$-1< \cos\theta_{c}<-0.8$~   & ~$75.1 \pm 12.3$~ \\
      ~$-0.8< \cos\theta_{h}<-0.6$~  & ~$63.9 \pm 11.3$~ &  ~$-0.8< \cos\theta_{c}<-0.6$~ & ~$68.2 \pm 11.6$~ \\
      ~$-0.6< \cos\theta_{h}<-0.4$~  & ~$68.9 \pm 11.7$~ & ~$-0.6< \cos\theta_{c}<-0.4$~  & ~$61.0 \pm 10.8$ ~ \\
      ~$-0.4< \cos\theta_{h}<-0.2$~  & ~$55.3 \pm 10.6$~ & ~$-0.4< \cos\theta_{c}<-0.2$~  & ~$33.9 \pm 9.0$ ~ \\
      $-0.2< \cos\theta_{h}< 0$~~~~~~& ~$57.5 \pm 11.1$~ & $-0.2< \cos\theta_{c}<0$~~~~~ & ~$37.0 \pm 9.6$ ~ \\
      ~~~$0< \cos\theta_{h}<0.2$~     & ~$90.2 \pm 12.0$~ & ~~~$0< \cos\theta_{c}<0.2$~       & ~$33.9 \pm 8.0$~ \\
      ~$0.2< \cos\theta_{h}<0.4$~    & ~$72.6 \pm 11.6$~ & ~$0.2< \cos\theta_{c}<0.4$~    & ~$37.7 \pm 9.8$~ \\
      ~$0.4< \cos\theta_{h}<0.6$~    & ~$53.3 \pm 10.1 $~ & ~$0.4< \cos\theta_{c}<0.6$~    & ~$48.2 \pm 10.1$~ \\
      ~$0.6< \cos\theta_{h}<0.8$~    & ~$50.6 \pm 9.8$~ & ~$0.6< \cos\theta_{c}<0.8$~    & ~$86.3 \pm 13.2$~ \\
      ~$0.8< \cos\theta_{h}<1$~~~   & ~$51.3 \pm 9.5$~ & ~$0.8< \cos\theta_{c}<1$~~~   & ~$94.9 \pm 12.6$~ \\
      \hline\hline
    \end{tabular}
  \end{center}
\end{table}

\begin{table}[!htb]
  \begin{center}
     \caption{ Summary of the reconstruction efficiency of the decay chain $\Xicpipi$ determined from signal MC events for each $\cos\theta_{h}$ and $\cos\theta_{c}$ bin. The uncertainties are statistical.
     }
        \label{tbl:eff_angle}
        \renewcommand{\arraystretch}{1.1}
    \begin{tabular}{c|c|c|c}\hline\hline
      \multirow{2}{*}{$\cos\theta_{h}$~} &  \multirow{2}{*}{~Efficiency [\%]~} &  \multirow{2}{*}{$\cos\theta_{c}$~} & \multirow{2}{*}{~Efficiency [\%]~}\\
      &~& \\
      \hline
      ~$-1< \cos\theta_{h}<-0.8$~       & ~{1.616 $\pm$ 0.001}~ &  ~$-1< \cos\theta_{c}<-0.8$~    & ~{2.537 $\pm$ 0.001}~ \\
      ~$-0.8< \cos\theta_{h}<-0.6$~	& ~{2.275 $\pm$ 0.001}~ &  ~$-0.8< \cos\theta_{c}<-0.6$~ & ~{2.529 $\pm$ 0.001}~ \\
      ~$-0.6< \cos\theta_{h}<-0.4$~	& ~{2.522 $\pm$ 0.001}~ & ~$-0.6< \cos\theta_{c}<-0.4$~  & ~{2.486 $\pm$ 0.001}~ \\
      ~$-0.4< \cos\theta_{h}<-0.2$~     & ~{2.636 $\pm$ 0.001}~ & ~$-0.4< \cos\theta_{c}<-0.2$~  & ~{2.467 $\pm$ 0.001}~ \\
      ~$-0.2< \cos\theta_{h}< 0$~       & ~{2.679 $\pm$ 0.001}~ & ~$-0.2< \cos\theta_{c}<0$~      & ~{2.451 $\pm$ 0.001}~ \\
      ~$0< \cos\theta_{h}<0.2$~         & ~{2.694 $\pm$ 0.001}~ & ~$0< \cos\theta_{c}<0.2$~       & ~{2.446 $\pm$ 0.001}~ \\
      ~$0.2< \cos\theta_{h}<0.4$~       & ~{2.660 $\pm$ 0.001}~ & ~$0.2< \cos\theta_{c}<0.4$~    & ~{2.439 $\pm$ 0.001}~ \\
      ~$0.4< \cos\theta_{h}<0.6$~        & ~{2.613 $\pm$ 0.001}~ & ~$0.4< \cos\theta_{c}<0.6$~    & ~{2.436 $\pm$ 0.001}~ \\
      ~$0.6< \cos\theta_{h}<0.8$~        & ~{2.546 $\pm$ 0.001}~ & ~$0.6< \cos\theta_{c}<0.8$~    & ~{2.441 $\pm$ 0.001}~ \\
      ~$0.8< \cos\theta_{h}<1$~          & ~{2.447 $\pm$ 0.001}~ & ~$0.8< \cos\theta_{c}<1$~       & ~{2.456 $\pm$ 0.001}~ \\
      \hline\hline
	

    \end{tabular}
  \end{center}
\end{table}
%
%
The yield of $\Xic$ is obtained by fitting the $\xicspi$ and $\xicppi$ invariant-mass distributions with all the phase space integrated for the $\xics$ and $\xicp$ signal regions.
The yield of $\Xic$ is $577\pm34$ in the $\xicspi$ decay mode and $201\pm33$ in the $\xicppi$ decay mode.
The reconstruction efficiency of $\Xic$ is determined from signal MC events, as shown in Table \ref{tbl:eff}.
The yields of $\xic$ ground states are obtained by fitting the invariant-mass distribution of $\xic$ candidates and reconstruction efficiency is determined from signal MC events.
The yield and reconstruction efficiency of the $\xic$ ground states are shown in Table \ref{tbl:yeff_gs}.
\begin{table}[!htb]
\begin{center}
\caption{ Summary of the reconstruction efficiency of $\Xic$ determined from signal MC events with all phase space integrated for the $\xics$ and $\xicp$ signal regions. The uncertainties are statistical.
     }
        \label{tbl:eff}
 \renewcommand{\arraystretch}{1.3}
    \begin{tabular}{c|c}\hline\hline
	~~Decay channel~~ & ~~ Efficiency [\%]~~ \\ \hline
	$\Xicspi$ with $\xic^+\to\Xi^-\pi^+\pi^+$ & ~$2.460\pm0.002$~ \\ 
	$\Xicppi$ with $\xic^0\to\Xi^-\pi^+$		& ~$2.136\pm0.002$~  \\
	$\Xicppi$ with $\xic^0\to\Omega^-K^+$	& ~$2.263\pm0.002$~  \\
	\hline\hline
   \end{tabular}
\end{center}
\end{table}

\begin{table}[!htb]
\begin{center}
\caption{ Summary of the yield and reconstruction efficiency of $\xic$ ground states. The yields are obtained by fitting the invariant-mass distribution of $\xic$ candidates and reconstruction efficiency is determined from signal MC events. The uncertainties are statistical.
     }
        \label{tbl:yeff_gs}
 \renewcommand{\arraystretch}{1.3}
    \begin{tabular}{c|cc}\hline\hline
	~~Decay channel~~ & ~~Yield ~~ & ~~ Efficiency [\%]~~ \\ \hline
	$\xic^+\to\Xi^-\pi^+\pi^+$ & ~$49627\pm268$~ & $10.52\pm0.01$ \\ 
	$\xic^0\to\Xi^-\pi^+$		& ~$36220\pm231$~ & $13.22\pm0.01$ \\
	$\xic^0\to\Omega^-K^+$	& ~$5307\pm78$~ & $11.32\pm0.01$ \\
	\hline\hline
   \end{tabular}
\end{center}
\end{table}

\end{document}